\documentclass[12pt,a4paper]{article}
\usepackage[latin1]{inputenc}
\usepackage{amsmath}
\usepackage{amsfonts}
\usepackage{amssymb}
\usepackage[left=2cm,right=2cm,top=2cm,bottom=2cm]{geometry}
\usepackage{tabularx}
\usepackage{subfig}
\usepackage{wrapfig}
\usepackage{tabularx}
\usepackage{sidecap}
\usepackage{setspace}
\usepackage{graphicx}
\usepackage{multirow}
\usepackage{multicol}
\usepackage{cite}
\RequirePackage[colorlinks,citecolor=blue,urlcolor=blue,linkcolor=blue]{hyperref}
\begin{document}
\title{A STUDY OF STRUCTURE FUNCTIONS WITH THE DGLAP EQUATIONS AT SMALL $x$ WITH $O(x)$ AND $O(x ^2 )$ .}
\author{Luxmi Machahari$^{1,a}$ , D. K. Choudhury$^{1,2,b}$ and P.K.Sahariah$^{3,c}$
\\ \\ $^1$Department of Physics, Gauhati University, Guwahati-781 014, India. \\$^2$Physics Academy of North-East, Guwahati-781 014, India.\\
$^3$Department of Physics,Cotton College,Guwahati 781001, India \\ $^a$email:luxmimachahari@gmail.com \\ $^b$email:dkc$\_$phys@yahoo.co.in \\ $^c$email:pksahariah@gmail.com}
\date{}
\maketitle
\doublespacing
\begin{abstract}
We obtain a pair of second order differential equations in two variables $x$ and $t$ from the coupled DGLAP QCD evolution equations at small $x$ using the standard Taylor series expansion method.To that end we keep terms upto $O(x^2 )$.We use the standard assumption about the relationship between the singlet Structure Function and the gluon distributions available in current literature. We solve the taylor approximated $O(x)$ DGLAP equations by Lagranges auxiliary method and $O(x^2)$ equation by Method of Separation of Variables and then show that the two solutions obtained in each for $O(x)$ and $O(x^2)$ are not identical in general.Analysis of the results obtained are done in the range of the recent HERA data.
\end{abstract}

\section{Introduction}
\label{intro}
\label{A}
The parton distribution function of the nucleon is an essential ingredient in any high energy physics experiments.The distribution functions which are functions of the Bjorken variable $x$ and the four momentum tranfer squared $Q^2$ between the lepton and the hadron in DIS experiment, is not calculable in theory but its evolution in $Q^2$ can be predicted by perturbative QCD method. The DGLAP equations \cite{gl,l,d,ap,esw} are the essential tools to calculate the evolution of the parton distributions and the structure functions of the nucleon and nuclei. These equations are integro-differential equations whose complexity rises as we move to higher and higher order. These equations are normally solved numerically taking into  account the world data on lepton hadron scattering \cite{h1}. But alternatively the evolution equations can also be solved analytically under certain assumptions valid to be at low $x$ \cite{DK2,DK1,DK4,DK6}. 
In reference \cite{DK2,DK1,DK4,DK6} several authors make a Taylor approximation and convert the integro differential DGLAP equations into first order partial differential equations in both $x$ and $t$ where $x$ is Bjorken variable and $t=$log$(\frac{Q^2}{\Lambda^2})$ and \ $\Lambda$ as QCD scale parameter.As a second step, an ansatz relating gluon and singlet structure Function is assumed \cite{DK4} to make them analytically solvable. 
In the present paper we rederive the solutions of the coupled DGLAP equations of $O(x)$ in LO at low $x$ and show that with such an ansatz the two equations in the coupled set lead to two different non-unique solutions by Lagranges Auxiliary method, 
a feature overlooked in Ref\cite{DK2,DK1,DK4,DK6}.Similarly we derive the solutions of the coupled DGLAP equations of $O(x^2)$ in LO at low $x$ by a very basic method of Separation of Variables and using the same ansatz, the coupled set lead to again two non-unique solutions. 


In sections \ref{B}, we outline a brief formalism,section \ref{C} shows the results and the conclusions are highlighted in section \ref{E}.
This paper reports a set of Taylor approximated $O(x)$ and $O(x^2)$ DGLAP Ref.\cite{gl,l,d,ap} equations at small $x$ using the Lagranges method and the Method of seperation of Variables respectively 
.
\section{Formalism}
\label{B}
The coupled DGLAP equations  for quark and gluon parton distribution function defined as $q(\textit{x}, \textit{t})$ and $g(\textit{x}, \textit{t})$ at leading order (LO)  are given in Ref.\cite{esw}
\begin{equation}
\label{E1}
t\frac{\partial}{\partial t}q_i(x,t) = \frac{\alpha_{s}(t)}{2\pi}\sum_{q_i\overline{q}_i}\int_x^{1}\dfrac{dz}{z}[ C_F\left\lbrace \frac{(1+z^2)}{(1-z)_+}+\frac{3}{2}\delta(1-z)\right\rbrace q_i(\frac{x}{z},t)+T_R\left\lbrace z^2+(1-z)^2\right\rbrace g(\frac{x}{z},t)] 
\end{equation}

\begin{multline}
\label{E2}
t\frac{\partial}{\partial t}g(x,t)=\frac{\alpha_{s}(t)}{2\pi}\sum_{q_i\overline{q}_i}\int_x^{1}\frac{dz}{z}[C_{F}\left\lbrace\frac{1+(1-z)^2}{z} \right\rbrace q_i(\frac{x}{z},t)+\lbrace2C_A\left( \frac{z}{(1-z)_+}\frac{1-z}{z}+z(1-z)\right) \\
+\delta(1-z)\frac{11C_{A}-4N_fT_R}{6}\rbrace g(\frac{x}{z},t)],
\end{multline}
where,
i is any quark flavor , \ $t=\log(\frac{Q^2}{\Lambda^2})$ , \ $Q^2$ is four momentum transfer squared with\ $\Lambda$ as QCD scale parameter.\ $\alpha_{s}$ is the strong coupling constant which is given in LO by $\alpha_s(Q^2)=\frac{12\pi}{(33-2N_f)log(Q^2/\Lambda^2)}$ \\ The other constants are $T_{R}=\frac{1}{2} ,\ \ \ C_{A}=3, \ \ \ C_{F}=\frac{4}{3}$ and\ $N_{f}$ is the number of quark Flavours.\\The subindex $'+'$ in the term $\frac{1}{(1-z)_+}$ is a plus distribution which indicates the cancellation of the divergence that is appearing at $z=1$ through 
$$\int_0^1dz\frac{f(z)}{(1-z)_+}=\int_0^1dz\frac{f(z)-f(1)}{1-z}$$ \ \ where, $f(z)$ is any arbitrary function.\\
 We reduce this integrodifferential equation to a calculable partial differential equation by defining a new variable $u=1-z$ and expanding $\frac{1}{z}$ as\cite{DK1}
\begin{equation}
\label{E3}
\frac{1}{z}=\frac{1}{u}=\sum_{r=0}^\infty u^r=1+\sum_{r=1}^\infty u^r
\end{equation}

Using the Taylor series upto the  second order O$(\textit{x}^2)$ for small \textit{x} for quarks and gluon distributions $q_i(\frac{x}{z})$ and $g(\frac{x}{z})$ can be written respectively as

\begin{equation}
\label{E4}
q_i(\frac{x}{z},t)=q_i(x,t)+x\sum_{r=1}^\infty u^r\frac{\partial}{\partial x}q_i(x,t)+\frac{x^2}{2!}(\sum_{r=1}^{\infty}u^r)^2\frac{\partial ^2}{\partial x^2}q_i(x,t)
\end{equation}

\begin{equation}
\label{E5}
g(\frac{x}{z},t)=g(x,t)+x\sum_{r=1}^\infty u^r\frac{\partial}{\partial x}g(x,t)+\frac{x^2}{2!}(\sum_{r=1}^{\infty}u^r)^2\frac{\partial ^2}{\partial x^2}g(x,t)
\end{equation}

Using equations (\ref{E4})and (\ref{E5}) in eqn.(\ref{E1}) the Taylor approximated form of the coupled DGLAP equations are:
\begin{multline}
\label{E6}
t\frac{\partial}{\partial t}q_i(x,t) = \frac{\alpha_{s}(t)}{2\pi}\sum_{q_i\overline{q}_i}\int_x^{1}\dfrac{dz}{z}[ C_F\left\lbrace \frac{(1+z^2)}{(1-z)_+}+\frac{3}{2}\delta(1-z)\right\rbrace \lbrace q_i(x,t)+x\sum_{r=1}^\infty u^r\frac{\partial}{\partial x}q_i(x,t)\\+\frac{x^2}{2!}(\sum_{r=1}^{\infty}u^r)^2\frac{\partial ^2}{\partial x^2}q_i(x,t)\rbrace +T_R\left\lbrace z^2+(1-z)^2\right\rbrace \lbrace g(x,t)+x\sum_{r=1}^\infty u^r\frac{\partial}{\partial x}g(x,t)\\+\frac{x^2}{2!}(\sum_{r=1}^{\infty}u^r)^2\frac{\partial ^2}{\partial x^2}g(x,t)\rbrace] 
\end{multline}

Similarly equation(\ref{E2}) can be written as
\begin{multline}
\label{E7}
t\frac{\partial}{\partial t}g(x,t)=\frac{\alpha_{s}(t)}{2\pi}\sum_{q_i\overline{q}_i}\int_x^{1}\frac{dz}{z}[C_{F}\left\lbrace\frac{1+(1-z)^2}{z} \right\rbrace\lbrace q_i(x,t)+x\sum_{r=1}^\infty u^r\frac{\partial}{\partial x}q_i(x,t)+\frac{x^2}{2!}(\sum_{r=1}^{\infty}u^r)^2\frac{\partial ^2}{\partial x^2}q_i(x,t)\rbrace\\+\lbrace 2C_A\left( \frac{z}{(1-z)_+}+\frac{1-z}{z}+z(1-z)\right)+\delta(1-z)\frac{11C_{A}-4N_fT_R}{6}\rbrace\lbrace g(x,t)\\+x\sum_{r=1}^\infty u^r\frac{\partial}{\partial x}g(x,t)+\frac{x^2}{2!}(\sum_{r=1}^{\infty}u^r)^2\frac{\partial ^2}{\partial x^2}g(x,t)\rbrace]
\end{multline}
Equations (\ref{E6}) and (\ref{E7}) can be written in compact form as:
\begin{equation}
\label{E8}
t\frac{\partial}{\partial t}q_i(x,t) = \frac{\alpha_{s}(t)}{2\pi}\sum_{q_i\overline{q}_i}[\lbrace J_1(x)+xJ_2(x)\frac{\partial}{\partial x}+x^2J_3(x)\frac{\partial ^2}{\partial x^2}\rbrace q_i(x,t)+\lbrace J_4(x)+xJ_5(x)\frac{\partial}{\partial x}+x^2J_6(x)\frac{\partial ^2}{\partial x^2}\rbrace g(x,t)]
\end{equation}
\begin{equation}
\label{E9}
t\frac{\partial}{\partial t}g(x,t) = \frac{\alpha_{s}(t)}{2\pi}\sum_{q_i\overline{q}_i}[\lbrace J_7(x)+xJ_8(x)\frac{\partial}{\partial x}+x^2J_9(x)\frac{\partial ^2}{\partial x^2}\rbrace q_i(x,t)+\lbrace J_{10}(x)+xJ_{11}(x)\frac{\partial}{\partial x}+x^2J_{12}(x)\frac{\partial ^2}{\partial x^2}\rbrace g(x,t)]
\end{equation}\\The expressions for $J_1,J_2,J_3,........J_{12}$ are as given in the Appendix A.\\
Neglecting the $O(x^2)$ terms in the equations (\ref{E8}) and (\ref{E9}), the corresponding DGLAP equations in $O(x)$ are as follows
\begin{equation}
\label{E8'}
t\frac{\partial}{\partial t}q_i(x,t) = \frac{\alpha_{s}(t)}{2\pi}\sum_{q_i\overline{q}_i}[\lbrace J_1(x)+xJ_2(x)\frac{\partial}{\partial x}\rbrace q_i(x,t)+\lbrace J_4(x)+xJ_5(x)\frac{\partial}{\partial x}\rbrace g(x,t)]
\end{equation}
\begin{equation}
\label{E9'}
t\frac{\partial}{\partial t}g(x,t) = \frac{\alpha_{s}(t)}{2\pi}\sum_{q_i\overline{q}_i}[\lbrace J_7(x)+xJ_8(x)\frac{\partial}{\partial x}\rbrace q_i(x,t)+\lbrace J_{10}(x)+xJ_{11}(x)\frac{\partial}{\partial x}\rbrace g(x,t)]
\end{equation}
To obtain the qualitative feature of our work, at the quark level,for a single quark flavor, we use the relation:
\begin{equation}
\label{E18'}
g(x,t)=C(t)q(x,t)
\end{equation}
Where, $C(t)$ is an $x$ independent and $t$ dependent function compatible with the analysis of Lopez and Yndurain \cite{ly}.
Using eqn.(\ref{E18'}) in equations(\ref{E8}) and (\ref{E9}) and also in equations(\ref{E8'}) and (\ref{E9'}) and  we obtain the following pair of $1^{st}$ and $2^{nd}$ order differential equations respectively as follows:
\begin{multline}
\label{E23i}
t\frac{\partial}{\partial t}q(x,t) = \frac{\alpha_{s}(t)}{2\pi}[J_1(x)+J_4(x)C(t)]q(x,t)+\frac{\alpha_{s}(t)}{2\pi}x[J_2(x)+J_5(x)C(t)]\frac{\partial}{\partial x}q(x,t)
\end{multline}

\begin{multline}
\label{E24i}
t\frac{\partial }{\partial t}(C(t)q(x,t)) = \frac{\alpha_{s}(t)}{2\pi}[J_7(x)+J_{10}(x)C(t)]q(x,t)+\frac{\alpha_{s}(t)}{2\pi}x[J_8(x)+J_{11}(x)C(t)]\frac{\partial}{\partial x}q(x,t)
\end{multline}

and 
\begin{multline}
\label{E23'}
t\frac{\partial}{\partial t}q(x,t) = \frac{\alpha_{s}(t)}{2\pi}[J_1(x)+J_4(x)C(t)]q(x,t)+\frac{\alpha_{s}(t)}{2\pi}x[J_2(x)+J_5(x)C(t)]\frac{\partial}{\partial x}q(x,t)+\frac{\alpha_{s}(t)}{2\pi}x^2[J_3(x)\\+J_6(x)C(t)]\frac{\partial ^2}{\partial x^2}q(x,t)
\end{multline}

\begin{multline}
\label{E24'}
t\frac{\partial }{\partial t}(C(t)q(x,t)) = \frac{\alpha_{s}(t)}{2\pi}[J_7(x)+J_{10}(x)C(t)]q(x,t)+\frac{\alpha_{s}(t)}{2\pi}x[J_8(x)+J_{11}(x)C(t)]\frac{\partial}{\partial x}q(x,t)\\+\frac{\alpha_{s}(t)}{2\pi}x^2[J_9(x)+J_{12}(x)C(t)]\frac{\partial ^2}{\partial x^2}q(x,t)
\end{multline}
equations (\ref{E23'}) and (\ref{E24'}) are solved by using the standard Lagranges Method \cite{s}.To apply Lagrange's Method, we put the two above equations (\ref{E23i}) and (\ref{E24i}) in the form as
\begin{equation}
\label{E21}
Q_1(x,t)\frac{\partial}{\partial t}q(x,t)+P_1(x,t)\frac{\partial}{\partial x}q(x,t)=R_1(x,t)
\end{equation}
\begin{equation}
\label{E22}
Q_2(x,t)\frac{\partial}{\partial t}q(x,t)+P_2(x,t)\frac{\partial}{\partial x}q(x,t)=R_2(x,t)
\end{equation}
Where, $Q_1(x,t),P_1(x,t),R_1(x,t),Q_2(x,t),P_2(x,t),R_2(x,t)$ are obtained by comparing equation(\ref{E21}) with equation(\ref{E23i}) and equation(\ref{E22}) with equation(\ref{E24i}) given in Appendix B.

The general solution of equation (\ref{E21}) and (\ref{E22}) is obtained by solving the following auxillary system of ordinary differential equations.
\begin{equation}
\frac{dx}{P_m(x,t)}=\frac{dt}{Q_m(x,t)}=\frac{dq(x,t)}{R_m(x,t)}
\end{equation}
Where, m goes from 1 to 2 referring respectively the two equations (\ref{E21}) and (\ref{E22}) and\\
If $u_m(x,t,q(x,t))=C_m$ and \ $v_m(x,t,q(x,t))=D_m$  are two independent solutions of the above equations (\ref{E21}) and (\ref{E22}),then in general their solutions are
\begin{equation}
\label{E23}
f_m(u_m,v_m)=0
\end{equation}
where, $f_m$ are arbitrary functions of $u_m$ and $v_m$.\\
Solving the auxillary equations, we get
\begin{eqnarray}
\label{E24}
u_1(x,t,q(x,t))=tX_1(x,t) \\
\label{E24a}
v_1(x,t,q(x,t))=q(x,t)Y_1(x,t) 
\end{eqnarray}

And, 
\begin{eqnarray}
\label{E25}
u_2(x,t,q(x,t))=X_2(x,t) \\
\label{E25a}
v_2(x,t,q(x,t))=q(x,t)Y_2(x,t)
\end{eqnarray}
Where $X_1(x,t),X_2(x,t),Y_1(x,t),Y_2(x,t)$ are in Appendix B.
The simplest possibility of finding a unique expression for $q(x,t)$ to satisfy the general solution (\ref{E23}) is the linear combination of $u_m$ and $v_m$ in $q(x,t)$ such as,
\begin{equation}
\label{E26}
u_m+\alpha_m v_m=\beta_m
\end{equation}
Where,\ $\alpha_m$ and $\beta_m$ are the quantities to be determined from boundary conditions on $q(x,t)$.
Considering $m=1$ and putting the expressions for $u_1$ and $v_1$ as in equations (\ref{E24}) and (\ref{E24a}) in the above equation(\ref{E26}) we get the corresponding solution as
\begin{equation}
\label{E27}
q^I(x,t)=\frac{1}{\alpha_1}\left[\frac{\beta_1}{Y_1(x,t)}-t\frac{X_1(x,t)}{Y_1(x,t)} \right] 
\end{equation}
Similarly for m=2 we obtain
\begin{equation}
\label{E28}
q^{II}(x,t)=\frac{1}{\alpha_2}\left[\frac{\beta_2}{Y_2(x,t)}-\frac{X_2(x,t)}{Y_2(x,t)} \right]
\end{equation}
where $\alpha_1,\alpha_2,\beta_1$ and $\beta_2$ are four unknown parameters.The algebraic structures show that (\ref{E27})$\neq$(\ref{E28}).\\

To obtain a corresponding information for second order we use the method of Separation of Variables \cite{ab} instead,which means all Parton Distribution Functions are factorizable in $x$ and $t$.Under this assumption we can write 
\begin{equation}
q^I(x,t)=X(x)T(t)
\end{equation}
 
\begin{equation} 
 q^{II}(x,t)=\hat{X}(x)\hat{T}(t)
\end{equation} 
  To go further we have to put some information of $C(t)$.Assuming $C(t)=k$(constant) \cite{DK4},We obtain the solutions for $T(t)$ and  $\hat{T}(t)$ as 
\begin{equation}  
  T(t)=\exp[-s^2\frac{6}{31}(\frac{1}{t}-\frac{1}{t_0})]
\end{equation}  
 and 
\begin{equation}   
   \hat{T}(t)=\exp[-\hat{s}^2\frac{6}{31k}(\frac{1}{t}-\frac{1}{t_0})]
\end{equation}
where $s^2$ and $\hat{s}^2$ are positive separation constants and for any initial scale at $Q^2=Q_0^2$ , $t_0=log(\frac{Q_0^2}{\Lambda^2})$.In obtaining both the expressions we used at any initial scale at $Q^2=Q_0^2$ (i.e  at $t=t_0$),the input distribution is a function of $x$ only.Hence $q(x,t_0)=X(x)$ or $\hat{X}(x)$ and so $T(t_0)=\hat{T}(t_0)=1$. 
Depending on the nature of equations Hyperbolic,Parabolic and Elliptic \cite{s}.$X(x)$ has three solutions:\\
a) $X(x)=\left[ c_1e^{m_1.x}+c_2e^{m_2.x}\right]$,    Hyperbolic\\
b) $X(x)=\left[ c_1e^{m.x}+c_2xe^{m.x}\right]$,     Parabolic\\
c) $X(x)=\left[ e^{\alpha x}(c_1 \cos \beta x+c_2 \sin \beta x)\right]$,   Elliptic

Consequently $q^I(x,t)$ have the following forms
\begin{equation}
\label{E29}
q^I(x,t)=\exp[-s^2\frac{6}{31}(\frac{1}{t}-\frac{1}{t_0})].\left[ c_1e^{m_1.x}+c_2e^{m_2.x}\right]
\end{equation}
\begin{equation}
\label{E30}
q^I(x,t)=\exp[-s^2\frac{6}{31}(\frac{1}{t}-\frac{1}{t_0})].\left[ c_1e^{m.x}+c_2xe^{m.x}\right]
\end{equation}
\begin{equation}
\label{E31}
q^I(x,t)=\exp[-s^2\frac{6}{31}(\frac{1}{t}-\frac{1}{t_0})].\left[ e^{\alpha x}(c_1 \cos \beta x+c_2 \sin \beta x)\right]
\end{equation}
and $q^{II}(x,t)$ have the forms
\begin{equation}
\label{E32}
q^{II}(x,t)=\exp[-\hat{s}^2\frac{6}{31k}(\frac{1}{t}-\frac{1}{t_0})].\left[ \hat{c}_1e^{\frac{-\hat{B}+ \sqrt{\hat{B}^2-4\hat{A}\hat{C}}}{2\hat{A}}.x}+\hat{c}_2e^{\frac{-\hat{B}- \sqrt{\hat{B}^2-4\hat{A}\hat{C}}}{2\hat{A}}.x}\right]
\end{equation}
\begin{equation}
\label{E33}
q^{II}(x,t)=\exp[-\hat{s}^2\frac{6}{31k}(\frac{1}{t}-\frac{1}{t_0})].\left[ \hat{c}_1e^{m.x}+\hat{c}_2xe^{m.x}\right]
\end{equation}
\begin{equation}
\label{E34}
q^{II}(x,t)=\exp[-\hat{s}^2\frac{6}{31k}(\frac{1}{t}-\frac{1}{t_0})].\left[ e^{\hat{\alpha} x}(\hat{c}_1 \cos \hat{\beta} x+\hat{c}_2 \sin \hat{\beta} x)\right]
\end{equation}
$c_1$, $c_2$,$\hat{c}_1$, and $\hat{c}_2$ are  arbitrary constants.
 $m_1$ and $m_2$ are defined 
as $\ m_1=\frac{-B+ \sqrt{B^2-4.A.H}}{2.A} ;\ \ m_2=\frac{-B- \sqrt{B^2-4.A.H}}{2.A}$ with
\begin{eqnarray}
\label{E32'}
A(x)=x^2[J_3(x)+J_6(x)k]\nonumber \\
B(x)=x[J_2(x)+J_5(x)k] \nonumber \\
H(x)=[J_1(x)+J_4(x)k]-s^2
\end{eqnarray} 
and
\begin{eqnarray}
\label{E46'}
\hat{A}(x)=x^2[J_9(x)+J_{12}(x)k]\nonumber \\
\hat{B}(x)=x[J_8(x)+J_{11}(x)k] \nonumber \\
\hat{H}(x)=[J_7(x)+J_{12}(x)k]-\hat{s}^2
\end{eqnarray}
To obtain the above forms (\ref{E29}),(\ref{E30}),(\ref{E31}),(\ref{E32}),(\ref{E33}),(\ref{E34}) analytically we make further assumption besides method of Separation of Variables such as  $A(x)\simeq A, \ B(x,t)\simeq B, \ H(x,t)\simeq H$ which means that $ A, B, H$ has a slow $x$ dependence.Thus here too from independent algebraic structures  $(\ref{E29})\neq(\ref{E32}),(\ref{E30})\neq(\ref{E33}),(\ref{E31})\neq(\ref{E34})$.
 An observation of above equations shows that algebraic structures of $q^I(x,t)$ and $q^{II}(x,t)$ for each are different i.e $q^I(x,t)\neq q^{II}(x,t)$.
 \\To conclude, we obtain two inequivalent quark distribution for small $x$ DGLAP equations taking both the $O(x)$ and $O(x^2)$ terms.
\section{Results}
\label{C}
In this section we show our results of $O(x)$ and $O(x^2)$ graphically.
\subsection{For $O(x)$:}
We can write from equations (\ref{E27}) and (\ref{E28})
\begin{equation}
\label{E32a}
q^I(x,t)=a_1(x,k)[\beta_1-t e^{a_2(x,t,k)}]
\end{equation}
\begin{equation}
\label{E32b}
q^{II}(x,t)=a_3(x,k)[\beta_2-t e^{a_4(x,t,k)}]
\end{equation}
where $a_1(x,k),a_2(x,t,k),a_3(x,k),a_4(x,t,k)$ are given in Appendix C.\\For the initial scale $t=t_0$
\begin{equation}
\label{E32c}
q^I(x,t_0)=a_1(x,k)[\beta_1-t_0 e^{a_2(x,t_0,k)}]
\end{equation}
\begin{equation}
\label{E32d}
q^{II}(x,t_0)=a_3(x,k)[\beta_2-t_0 e^{a_4(x,t_0,k)}]
\end{equation}

Now we take the ratio of eqn.(\ref{E32a}) and (\ref{E32c}) as
\begin{equation}
\label{E32e}
\frac{q^I(x,t)}{q^I(x,t_0)}=\frac{[\beta_1-t e^{a_2(x,t,k)}]}{[\beta_1-t_0 e^{a_2(x,t_0,k)}]}
\end{equation}
ratio of eqn.(\ref{E32b}) and (\ref{E32d})
\begin{equation}
\label{E32f}
\frac{q^{II}(x,t)}{q^{II}(x,t_0)}=\frac{[\beta_2-t e^{a_4(x,t,k)}]}{[\beta_2-t_0 e^{a_4(x,t_0,k)}]}
\end{equation}

So,$$q^I(x,t)=q^I(x,t_0)\frac{q^I(x,t)}{q^I(x,t_0)}$$
\begin{equation}
\label{E32g}
\implies q^I(x,t)=q(x,t_0)\frac{[\beta_1-t e^{a_2(x,t,k)}]}{[\beta_1-t_0 e^{a_2(x,t_0,k)}]}
\end{equation}

Similarly,
\begin{equation}
\label{E32h}
 q^{II}(x,t)=q(x,t_0)\frac{[\beta_2-t e^{a_4(x,t,k)}]}{[\beta_2-t_0 e^{a_4(x,t_0,k)}]}
\end{equation}
where we assumed $q^I(x,t_0)=q^{II}(x,t_0)=q(x,t_0)$ (input distribution)\\
The above two equations (\ref{E32g}) and (\ref{E32h}) are the exact expressions for our model parton distribution function at $O(x)$.
To express it in terms of Singlet Structure Functions $F_2^S(x,t)=x\sum_i\left\lbrace q_i(x,t)+\bar{q}_i(x,t)\right\rbrace $  we can write the above two expressions in the limit
\begin{equation}
\label{E03} 
q_i(x,t)=\bar{q}_i(x,t)
\end{equation}
 as 
\begin{equation}
\label{E04} 
 F_2^{S(I)}(x,t)=2x\sum_i^{N_f}q_i^I(x,t)=2xN_fq_{i}^I(x,t)
\end{equation} 

\begin{equation}
\label{E05}  
 F_2^{S(II)}(x,t)=2x\sum_i^{N_f}q_i^{II}(x,t)=2xN_fq_{i}^{II}(x,t)
 \end{equation} 

Now we need to numerically determine the unknown terms $\beta_1,\beta_2,k$. In the ultra small $x$ limit we get 
\begin{equation}
\label{E06} 
x(J_2(x)+J_5(x)k)=\frac{4}{3}+\frac{k}{2}
\end{equation}
and 
\begin{equation}
\label{E07} 
x(J_8(x)+J_{11}(x)k)=(\frac{1}{x}-4)(\frac{4}{3}+3k)
\end{equation}
In this limit we obtain the expression for the following

\begin{equation}
\label{E08} 
a_2(x,t,k)=t\frac{33-2N_f}{6}\frac{x}{\frac{4}{3}+\frac{k}{2}}
\end{equation}
 and 
\begin{equation}
\label{E09}  
 a_4(x,t,k)=t\frac{33-2N_f}{6}\frac{1}{(\frac{4}{3}+3k)}(\frac{x^2}{2}+\frac{4x^3}{3})-\frac{1}{k}log(t)
 \end{equation}
In that ultra small $x$ limit (\ref{E32g}) and (\ref{E32h}) becomes
\begin{equation}
\label{E32i}
q^I(x,t)=q(x,t_0)\frac{\left[\beta_1-t \exp\left(t\frac{33-2N_f}{6}\frac{x}{\frac{4}{3}+\frac{k}{2}}\right)\right]}{\left[\beta_1-t_0 \exp\left(t_0\frac{33-2N_f}{6}\frac{x}{\frac{4}{3}+\frac{k}{2}}\right)\right]}
\end{equation}

\begin{equation}
\label{E32j}
q^{II}(x,t)=q(x,t_0)\frac{\left[\beta_2-t \exp\left(t\frac{33-2N_f}{6}\frac{1}{(\frac{4}{3}+3k)}(\frac{x^2}{2}+\frac{4x^3}{3})-\frac{1}{k}log(t)\right)\right]}{\left[\beta_2-t_0 \exp\left(t_0\frac{33-2N_f}{6}\frac{1}{(\frac{4}{3}+3k)}(\frac{x^2}{2}+\frac{4x^3}{3})-\frac{1}{k}log(t_0)\right)\right]}
\end{equation}
Now for any fixed $x$ and certain $Q^2$ we can determine $\beta_1$ and $k$ from equation(\ref{E32i}) and then we can determine $\beta_2$ from equation(\ref{E32j}) using the obtained values of $\beta_1$ and $k$ as given in Appendix D.\\ Taking the HERA input parameterization \cite{h1} and using the recent HERA data \cite{h1},we obtain 
\begin{equation}
\label{E32g1}
 q^I(x,t)=q(x,t_0)\frac{[-6.03-t e^{a_2(x,t,0.8)}]}{[-6.03-t_0 e^{a_2(x,t_0,0.8)}]}
\end{equation}

\begin{figure}[hbtp]
\begin{center}
\subfloat[]{\includegraphics[scale=0.902]{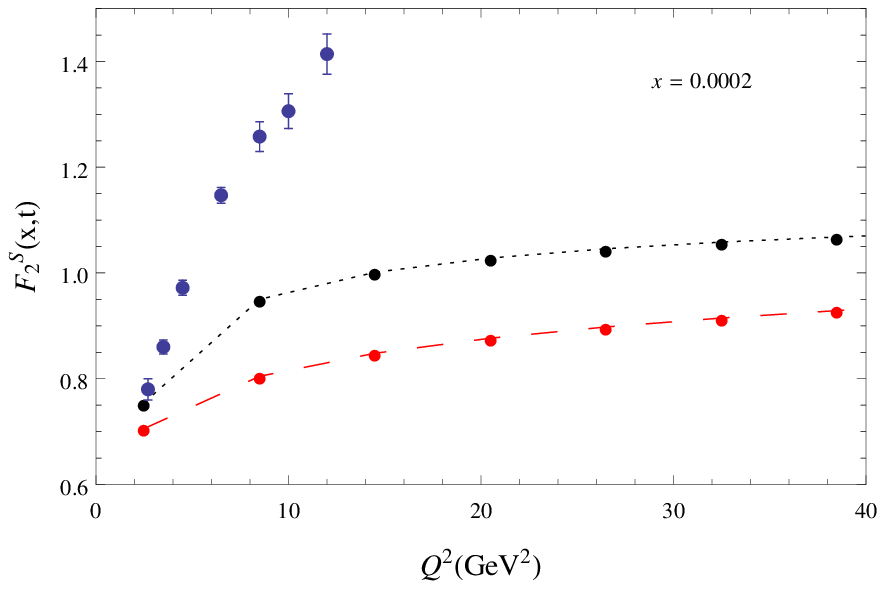}}\quad
\subfloat[]{\includegraphics[scale=0.902]{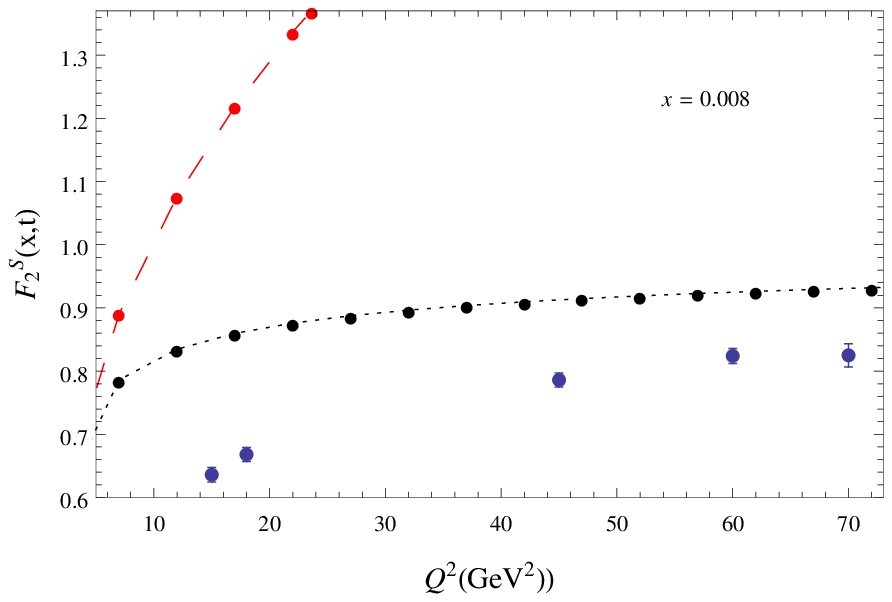}}\quad
\caption{Plot of Singlet Structure Function $F_2^S(x,t)$ as a function of $Q^2$ at different fixed $x$ with HERAPDF1.0 data.Here the red dashed-connected line represents our analytical result for $F_2^{S(I)}(x,t)$ and that of black dotted- connected line for $F_2^{S(II)}(x,t)$,obtained by Lagrange's Auxiliary method .}
\label{Fig1}
\end{center}
\end{figure}

And,
\begin{equation}
\label{E32h1}
 q^{II}(x,t)=q(x,t_0)\frac{[0.57-t e^{a_4(x,t,0.8)}]}{[0.57-t_0 e^{a_4(x,t_0,0.8)}]}
\end{equation}
In the expressions $a_2(x,t,0.8)$ and $a_4(x,t,0.8)$ there is an integration over $x$.So to obtain it numerically a lower and upper limit of $x$ is required.We take the lower limit of $x$ as $6\times10^{-7}$ \cite{h1} and upper limit as $0.2$.Our model is based on small $x$ so the highest limit we take here is 0.2 which is the plausible higher small $x$ limit.
We test graphically whether the structure functions $F_2^{S(I)}(x,t)$ and $F_2^{S(II)}(x,t)$ obtained from equations (\ref{E04}) and (\ref{E05}) using equations(\ref{E32g1}) and (\ref{E32h1})  respectively match well with the recent HERA data as in Fig.\ref{Fig1}.
We observe that the two solutions (\ref{E32g1}) and (\ref{E32h1}) have different $Q^2$ evolutions and second solution is much closer to the experimental data.

\subsection{For $O(x^2)$:}
We assume that for a very small $x$ ,we can take $\sin\beta x\simeq0$ in equation(\ref{E31})
\begin{equation}
\label{E35}
q^I(x,t)=\exp[-s^2\frac{6}{31}(\frac{1}{t}-\frac{1}{t_0})].\left[ e^{\alpha x}(c_1 \cos \beta x)\right]
\end{equation}
Likewise as in $O(x)$ 
\begin{equation}
\label{E36}
q^I(x,t)=q^I(x,t_0)\frac{q^I(x,t)}{q^I(x,t_0)}
\end{equation}
Now,
\begin{equation}
\label{E37}
\frac{q^I(x,t)}{q^I(x,t_0)}=\frac{\exp[-s^2\frac{6}{31}(\frac{1}{t}-\frac{1}{t_0})].\left[ e^{\alpha x}(c_1 \cos \beta x)\right]}{\left[ e^{\alpha x}(c_1 \cos \beta x)\right]}
\end{equation}
Putting in equation(\ref{E36}) we obtain,
\begin{equation}
\label{E38}
q^I(x,t)=q(x,t_0)\exp[-s^2\frac{6}{31}(\frac{1}{t}-\frac{1}{t_0})]
\end{equation}
Similarly for $q^{II}(x,t)$ of equation(\ref{E34}) when $\sin\hat{\beta}x\simeq0$
\begin{equation}
\label{E39}
q^{II}(x,t)=q(x,t_0)\exp[-\hat{s}^2\frac{6}{31k}(\frac{1}{t}-\frac{1}{t_0})]
\end{equation}
where in equations (\ref{E38}) and (\ref{E39}) we assumed $q^I(x,t_0)=q^{II}(x,t_0)=q(x,t_0)$ (input distribution).\\In terms of Singlet Structure Function we get,

\begin{eqnarray}
\label{E40}
F_2^{S(I)}(x,t)=F_2^{S}(x,t_0)\exp[-s^2\frac{6}{31}(\frac{1}{t}-\frac{1}{t_0})]
\end{eqnarray}

Again,
\begin{eqnarray}
\label{E41}
F_2^{S(II)}(x,t)=F_2^{S}(x,t_0)\exp[-\hat{s}^2\frac{6}{31k}(\frac{1}{t}-\frac{1}{t_0})]
\end{eqnarray}
Now we make a graphical interpretation of our obtained Singlet Structure Functions as a function of $Q^2$ for a certain $x$.We use the obtained value of $k$ as in $O(x)$ as 0.8 and we take separation constants $s^2=5$ and $\hat{s}^2=5$, as lower the value of separation constants makes the separation between the  $F_2^{S(I)}(x,t)$ and $F_2^{S(II)}(x,t)$ plots less.
\begin{figure}[hbtp]
\begin{center}
\subfloat[]{\includegraphics[scale=0.902]{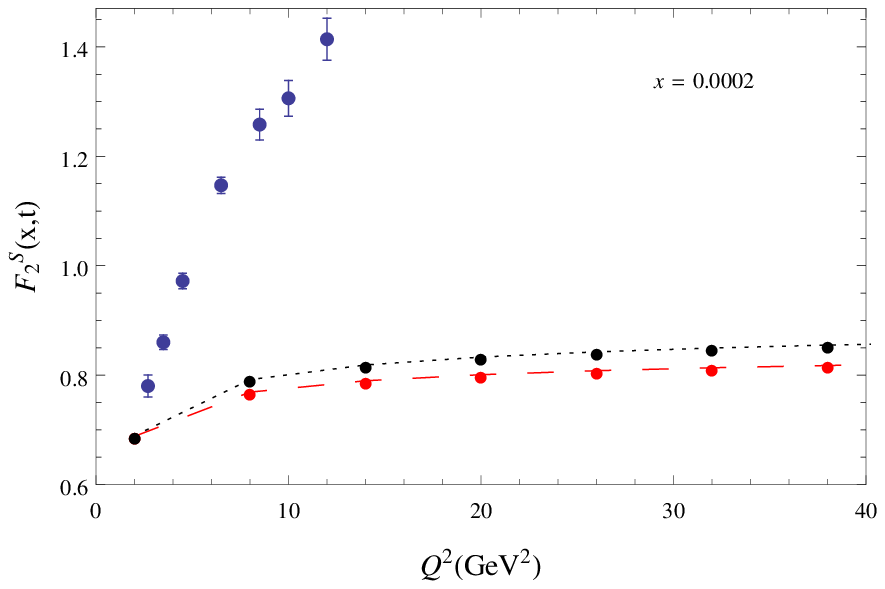}}\quad
\subfloat[]{\includegraphics[scale=0.902]{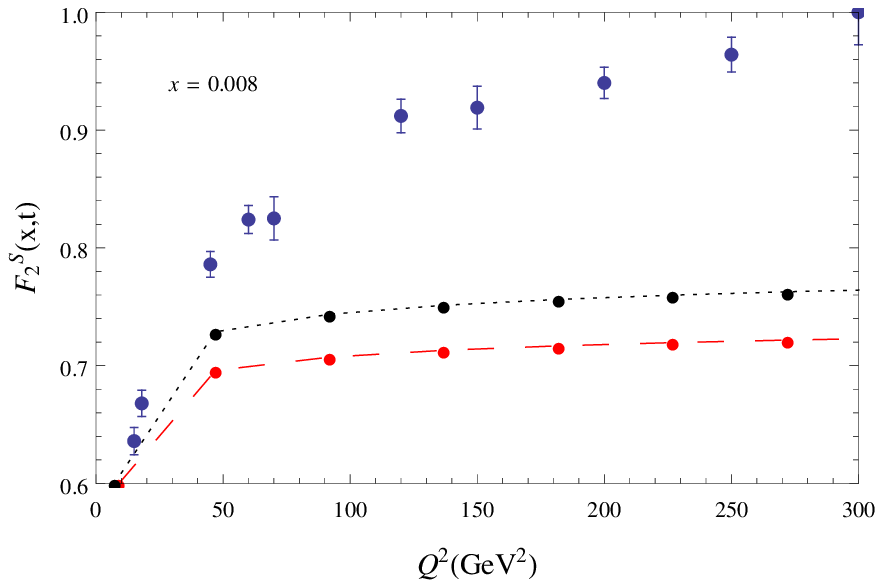}}\quad
\caption{Plot of Singlet Structure Function $F_2^S(x,t)$ as a function of $Q^2$ at different fixed $x$ with HERAPDF1.0 data.Here the red dashed-connected line represents our analytical result for $F_2^{S(I)}(x,t)$ and that of black dotted- connected line for $F_2^{S(II)}(x,t)$, obtained by the method of Separation of Variables. }
\label{Fig2}
\end{center}
\end{figure}
Fig.\ref{Fig2}(a) and fig.\ref{Fig2}(b) have the same feature as in fig.\ref{Fig1}(a) and fig.\ref{Fig1}(b) which shows the non identical nature of solutions and the second solution is closer to the data.

\section{Conclusion}
\label{E}
In this paper we have shown that the two solutions $q^I(x,t)$ and $q^{II}(x,t)$ obtained from the DGLAP equations (\ref{E23i}) and (\ref{E24i}) for $O(x)$ and from equations (\ref{E23'}) and (\ref{E24'}) for $O(x^2)$ are not identical with each other,a feature not studied in the earlier communications \cite{DK2,DK1,DK4,ne}.The graphical representations  also support this conclusion.While comparing with the recent HERA data, we find that our theoretical prediction of first solution overshoots the experimental data and the second one is much closer to it. 
Proper incorporation of flavour dependence in the equations (\ref{E18'}) and (\ref{E03}) might perhaps improve our result further significantly.Such study is currently under progress.

\section*{Acknowledgment}
One of the authors (L.M) acknowledges the Rajiv Gandhi National Fellowship, New Delhi for financial support.

\section*{Appendix A:}

$J_1(x)=\int_x^{1}\frac{dz}{z}C_{F}\left\lbrace\frac{(1+z^2)}{(1-z)_{+}}+\frac{3}{2}\delta(1-z)\right\rbrace \\
J_2(x)=\int_x^{1}\frac{dz}{z}\sum_{r=1}^\infty u^r C_{F}\left\lbrace\frac{(1+z^2)}{(1-z)_{+}}+\frac{3}{2}\delta(1-z)\right\rbrace \\
J_3(x)=\int_x^{1}\frac{dz}{z}\frac{(\sum_{r=1}^\infty u^r)^2}{2!} C_{F}\left\lbrace\frac{(1+z^2)}{(1-z)_{+}}+\frac{3}{2}\delta(1-z)\right\rbrace \\
 J_4(x)=\int_x^{1}\frac{dz}{z}T_R\left\lbrace z^2+(1-z)^2\right\rbrace \\
J_5(x)=\int_x^{1}\frac{dz}{z} T_R\left\lbrace z^2+(1-z)^2\right\rbrace\sum_{r=1}^\infty u^r\\
J_6(x)=\int_x^{1}\frac{dz}{z} T_R\left\lbrace z^2+(1-z)^2\right\rbrace\frac{(\sum_{r=1}^\infty u^r)^2}{2!}\\
J_7(x)=\int_x^{1}\frac{dz}{z}C_{F}\left\lbrace\frac{(1+(1-z)^2)}{z}\right\rbrace \\
J_8(x)=\int_x^{1}\frac{dz}{z}C_{F}\left\lbrace\frac{(1+(1-z)^2)}{z}\right\rbrace\sum_{r=1}^\infty u^r\\
J_9(x)=\int_x^{1}\frac{dz}{z}C_{F}\left\lbrace\frac{(1+(1-z)^2)}{z}\right\rbrace\frac{(\sum_{r=1}^\infty u^r)^2}{2!}\\
J_{10}(x)=\int_x^{1}\frac{dz}{z}\left\lbrace 2C_{A}\left( \frac{z}{(1-z)_+}+\frac{1-z}{z}+z(1-z)\right)+\delta(1-z)\frac{11C_A-4N_fT_R}{6}\right\rbrace \\
J_{11}(x)=\int_x^{1}\frac{dz}{z}\left\lbrace 2C_{A}\left( \frac{z}{(1-z)_+}+\frac{1-z}{z}+z(1-z)\right)+\delta(1-z)\frac{11C_A-4N_fT_R}{6}\right\rbrace\sum_{r=1}^\infty u^r\\
J_{12}(x)=\int_x^{1}\frac{dz}{z}\left\lbrace 2C_{A}\left( \frac{z}{(1-z)_+}+\frac{1-z}{z}+z(1-z)\right)+\delta(1-z)\frac{11C_A-4N_fT_R}{6}\right\rbrace\frac{(\sum_{r=1}^\infty u^r)^2}{2!}$

\section*{Appendix B:}
$X_1(x,t)=exp\left[-\int\frac{dx}{P_1(x,t)}\right]\\
Y_1(x,t)=exp\left[-\int\frac{dx}{P_1(x,t)}\hat{R}_1(x,t)\right]\\
X_2(x,t)=exp\left[-\int\frac{dx}{P_2(x,t)}-\int\frac{dt}{tC(t)}\right]\\
Y_2(x,t)=exp\left[-\int\frac{dx}{P_2(x,t)}\hat{R}_2(x,t)\right]\\
Q_1(x,t)=t \nonumber\\
P_1(x,t)=-\frac{\alpha_s(t)}{2\pi}x[J_2(x)+J_5(x)C(x,t)]\\
R_1(x,t)=\frac{\alpha_s(t)}{2\pi}[J_1(x)+J_4(x)C(x,t)]q(x,t)=\hat{R_1}(x,t)q(x,t)\\
Q_2(x,t)=tC(t) \\
P_2(x,t)=-\frac{\alpha_s(t)}{2\pi}x[J_8(x)+J_{11}(x)C(x,t)]\\
R_2(x,t)=[\frac{\alpha_s(t)}{2\pi}\left\lbrace J_7(x)+J_{10}(x)C(x,t)\right\rbrace -t\frac{\partial}{\partial t}C(x,t)] q(x,t)=\hat{R_2}(x,t)q(x,t)\\$

\section*{Appendix C:}
$$a_1(x,k)=\frac{1}{\alpha_1\exp\left( -\int\frac{\frac{\alpha_s(t)}{2\pi}(J_1(x)+J_4(x)k)}{-\frac{\alpha_s(t)}{2\pi}x(J_2(x)+J_5(x)k)}dx\right) }$$

$$a_2(x,k)=-\int\frac{dx}{-\frac{\alpha_s(t)}{2\pi}x(J_2(x)+J_5(x)k)}$$

$$a_3(x,k)=\frac{1}{\alpha_2\exp\left( -\int\frac{\frac{\alpha_s(t)}{2\pi}(J_7(x)+J_{10}(x)k)}{-\frac{\alpha_s(t)}{2\pi}x(J_8(x)+J_{11}(x)k)}dx\right) }$$

$$a_4(x,k)=-\int\frac{dx}{-\frac{\alpha_s(t)}{2\pi}x(J_8(x)+J_{11}(x)k)}-\frac{1}{k}log(t)$$

\section*{Appendix D:}
For different inputs $q^I(x,t)$ and $q^{II}(x,t)$ will evolve differently.For $q(x,t_0)$ we take the input parameterization of HERA group at $Q_0^2=1.9$ GeV$^2$ 
 as follows.
$$xu_v(x)=A_{u_v}x^{B_{u_v}}(1-x)^{C_{u_v}}(1+E_{u_v}x^2)$$
$$xd_v(x)=A_{d_v}x^{B_{d_v}}(1-x)^{C_{d_v}}$$
$$x\bar{U}(x)=A_{\bar{U}}x^{B_{\bar{U}}}(1-x)^{C_{\bar{U}}}$$
$$x\bar{D}(x)=A_{\bar{D}}x^{B_{\bar{D}}}(1-x)^{C_{\bar{D}}}$$ \\The values of the parameters $A_{u_v},B_{u_v},C_{u_v},E_{u_v}$ etc are as given in Appendix C Table\ref{Table1}.

If $xU,xD,x\bar{U}$ and $x\bar{D}$ denote the sums of up,down and their corresponding anti-quark distributions respectively and are related to quark distributions as:
$$xU=xu+xc$$ $$x\bar{U}=x\bar{u}+x\bar{c}$$ $$xD=xd+xs$$ $$x\bar{D}=x\bar{d}+x\bar{s}$$
$xs,xc$ are strange and charm quark distributions.
Assuming sea quarks and anti-quarks symmetry yields the valence quark distributions as $$xu_v=xU-x\bar{U}$$ $$xd_v=xD-x\bar{D}$$
Now the singlet structure function $F_2^S(x)$ for the three flavors $u,d,s$ is:
$$ F_2^S(x)=x\left\lbrace u(x)+\bar{u}(x)\right\rbrace+x\left\lbrace d(x)+\bar{d}(x)\right\rbrace+x\left\lbrace s(x)+\bar{s}(x)\right\rbrace  $$
Where $u(x)=u_v(x)+u_s(x)$ and so for $d(x)$ and $s(x)$.$u_v(x)$ and $u_s(x)$ are the $u$ valence and $u$ sea quark distributions respectively and we assume $u_s\approx d_s\approx s_s \approx \bar{u}_s\approx \bar{d}_s\approx \bar{s}_s$ for $x\rightarrow0$,
So the singlet structure function becomes $$F_2^S(x)=xu_v(x)+xd_v(x)+6xs_s(x)$$
$$\implies F_2^S(x)=xu_v(x)+xd_v(x)+3x\bar{D}(x)$$
We take two data points for a fixed $x$ and two different $Q^2$ 150 GeV$^2$ and 200 GeV$^2$ and their corresponding parton distribution data values \cite{h1}. \\For $x=0.800\times10^{-2}, Q^2=150$ GeV$^2$,$F_2^S(x,t)=0.919$ and $t_0=$log$\frac{Q_0^2}{\Lambda^2}$, $N_f=3$,$\Lambda^{(3)}=340$MeV, $F_2^S(x,t_0)=xq(x,t_0)=0.578639$ we obtain an equation interms of $\beta_1$ and $k$ from equation(\ref{E32i}) as
\begin{equation}
\label{E32k}
0.58821\beta_1 -4.44615 \times \exp(\frac{0.604686}{8+3k})+7.16825\times \exp(\frac{1.54834}{8+3k})=0
\end{equation}

Likewise for $x=0.800\times10^{-2}, Q^2=200$ GeV$^2$,$F_2^S(x,t)=0.940$, we obtain from equation(\ref{E32i}) as 
\begin{equation}
\label{E32l}
0.6245\beta_1 -4.54774 \times \exp(\frac{0.604686}{8+3k})+7.45594\times \exp(\frac{1.61048}{8+3k})=0
\end{equation}

From eqn(\ref{E32k}) we can write as $$\beta_1=7.55878\times \exp(\frac{0.604686}{8+3k})-12.1865\times \exp(\frac{1.54834}{8+3k})$$ and use it in the equation(\ref{E32l}) which yields as 
\begin{equation}
\label{E32m}
0.17\times\exp(\frac{0.60}{8+3k})-7.61\times\exp(\frac{1.54}{8+3k})+7.45\times\exp(\frac{1.61}{8+3k})=0
\end{equation}
To make our mathematics simpler we assume the expansion series of $e^x$ upto the first order of $x$ in  $e^x=1+\frac{x}{1!}+\frac{x^2}{2!}+...$ and incorporate in 
the above equation(\ref{E32m}) and we obtain the value of $k$ as 0.8 and henceforth we obtain the value of $\beta_1$ as -6.03 from equation(\ref{E32l}) and $\beta_2$ to be 0.57 from equation(\ref{E32j}) with $x=0.800\times10^{-2}, Q^2=150$ GeV$^2$,$F_2^S(x,t)=0.919$\\

\begin{table}[h]
\caption{Cental values of the HERAPDF1.0 parameters}
\label{Table1}
\begin{center}
\begin{tabularx}{12cm}{|X|X|X|X|X|}
\hline
 & $A$ & $B$ & $C$ & $E$  \\ \hline
 $xg$ & 6.8 & 0.22   & 9.0 & \\ \hline
 $xu_v$ & 3.7 & 0.67 & 4.7 & 9.7 \\ \hline
 $xd_v$ & 2.2 & 0.67 & 4.3 &\\ \hline
 $x\bar{U}$ & 0.113 & -0.165 & 2.6 &\\ \hline
 $x\bar{D}$ & 0.163 & -0.165 & 2.4 &\\ \hline
\end{tabularx}

\end{center}
\end{table}


\begin{thebibliography}{10}

\bibitem{gl} V. N. Gribov and L. N. Lipatov, Sov. J. Nucl. Phys., 15, 438 (1972);
\bibitem{l} L. N. Lipatov, Sov. J. Nucl. Phys., 20, 94 (1975);
\bibitem{d} Yu. L. Dokshitzer, Sov. Phys. JETP, 46, 641 (1977);
\bibitem{ap} G. Altarelli and G. Parisi, Nucl. Phys. B126, 298 (1977);
\bibitem{esw} QCD and collider Physics,R.K Ellis,W.J Stirling and B.R.Webber

\bibitem{h1}  H1 and ZEUS Collaborations (Aaron, F.D. et al.) JHEP 1001 (2010) 109 arXiv:0911.0884 [hep-ex] DESY-09-158 
\bibitem{DK2} D.K. Choudhury and J.K. Sarma, \textit{Pramana J. Phys.},Vol.38, No.5, May 1992, pp.481-489. 
\bibitem{DK4}J. K. Sarma, D. K. Choudhury and G. K. Medhi, \textit{Phys. Lett. B}403, 139 (1997).
\bibitem{DK1} Neelakshi N.K. Borah, D.K. Choudhury and P.K. Sahariah, \textit{Advances in High Energy Physics} Volume 2013 Article ID 829803, 10 pages.
\bibitem{s} I. Sneddon,(1957)p131; Elements of Partial Differential Equations,M.Graw Hill, NewYork
\bibitem{ab} A.B. Gupta,(2004)p381; Fundamentals of Mathematical Physics,Books and Allied (P) LTD,Kolkata;
\bibitem{DK6} D.K Choudhury Journal of Physics,Conference series 481(2014)012021
\bibitem{DK5} A. Jahan and D.K. Choudhury, \textit{Commun.Theor. Phys.} 61(2014) 654-658

\bibitem{ly}C. Lopez and F. J. Yndurain, \textit{Nucl. Phys. B}171, 231 (1980).
\bibitem{ne} D.K. Choudhury and Neelakshi N.K. Borah arXiv:1508.06041v3 [hep-ph] 14 May 2016 accepted for publication in Physical Review D.

 


\end{thebibliography}
\end{document}